\documentclass[12pt]{article}
\usepackage{graphicx}

\usepackage[T1]{fontenc}

\usepackage{siunitx}
\usepackage{url}

\newcommand\pubnumber{CMS-CR-2018/370}
\newcommand\pubdate{\today}

\def\institute{University of Bristol}

\def\Title#1{\begin{center} {\Large #1 } \end{center}}
\def\Author#1{\begin{center}{ \sc #1} \end{center}}
\def\Address#1{\begin{center}{ \it #1} \end{center}}

\newcommand\pubblock{\rightline{\begin{tabular}{l} \pubnumber\\
         \pubdate  \end{tabular}}}
\newenvironment{Abstract}{\begin{quotation}  }{\end{quotation}}
\newenvironment{Presented}{\begin{quotation} \begin{center} 
             PRESENTED AT\end{center}\bigskip 
      \begin{center}\begin{large}}{\end{large}\end{center} \end{quotation}}

\input econfmacros.tex

\begin{document}
\begin{titlepage}
\pubblock

\vfill
\Title{Top Quark Modelling and Tuning at CMS}
\vfill
\Author{Emyr Clement \\
on behalf of the CMS Collaboration}
\Address{\institute}
\vfill
\begin{Abstract}
Recent measurements dedicated to improving the understanding of modelling top quark pair (\ensuremath{{\text{t}\overline{\text{t}}}}\xspace) production at the LHC are summarised.  These measurements, performed with proton-proton collision data collected by the CMS detector at $\sqrt{s}=$ 13~TeV, probe the underlying event in \ensuremath{{\text{t}\overline{\text{t}}}}\xspace events, and use the abundance of jets in \ensuremath{{\text{t}\overline{\text{t}}}}\xspace events to study the substructure of jets.  A new set of tunes for \textsc{pythia 8}, and their performance with \ensuremath{{\text{t}\overline{\text{t}}}}\xspace data, are also discussed.
\end{Abstract}
\vfill
\begin{Presented}
$11^\mathrm{th}$ International Workshop on Top Quark Physics\\
Bad Neuenahr, Germany, September 16--21, 2018
\end{Presented}
\vfill
\end{titlepage}
\def\thefootnote{\fnsymbol{footnote}}
\setcounter{footnote}{0}

\section{Introduction}

Accurate modelling of top quark production at the LHC is a crucial aspect of top quark measurements.  This is particularly true as modelling uncertainties often dominate over statistical and experimental sources of uncertainty in our measurements, and become a limiting factor in the precision of the measurements.  The abundance of data collected by the CMS detector~\cite{CMS} allows extensive studies of top quark production to be performed with the aim of probing the accuracy of various aspects of Monte Carlo (MC) generators.

In these proceedings, two recent measurements, dedicated to improving the understanding of the modelling of top quark pair (\ttbar) production, are presented.  The predictions of a new set of tunes for the \PYTHIA parton shower (PS) generator~\cite{Pythia} are also presented, along with their performance when compared to top quark data.

\section{Simulation of \ttbar events}

In the two measurements presented here, the nominal predictions for \ttbar production are obtained using the \POWHEG (v2) matrix element (ME) generator~\cite{Powheg1,Powheg2,Powheg3,PowhegHVQ}, interfaced with \PYTHIA for the simulation of the PS. The ME calculations are at next-to-leading-order (NLO) accuracy, and \PYTHIA employs the \cuettune tune~\cite{CUETP8M2T4}.  This tune was derived to address a mismodelling of the jet multiplicity observed with the previous tune, \oldcuettune~\cite{CUETP8M1}.  The data are also compared with predictions obtained with other state-of-the-art ME and PS generators, such as \MG~\cite{MG}, \SHERPA~\cite{Sherpa}, and \HERWIGS~\cite{Herwig7_1,Herwig7_2}.

\section{The Underlying Event in \ttbar~events}
\label{sec:UE}
Measurements of the underlying event (UE) in the LHC environment are typically performed using minimum bias (MB) or DY events.  Using \ttbar~events to study the UE probes the universality of the UE modelling at higher scales, of around twice the top mass (\topmass).  The properties of the UE, namely the multiplicity, momentum flux, and UE shape, have been measured by CMS with 35.9\fbinv of proton-proton collision data at $\sqrts=13\TeV$~\cite{TOP17015}.  The measurement uses \ttbar events with two leptons (\epm or \mupm) and two {\bJet}s in the final state, which allows these products from the hard scatter, and those associated with pileup interactions, to be separated from the charged particles associated with the UE.  The measured properties of the UE are unfolded to particle level, to a phase space that mimics that used to select the events.

\begin{figure}[htb]
\centering
\includegraphics[width=0.42\textwidth]{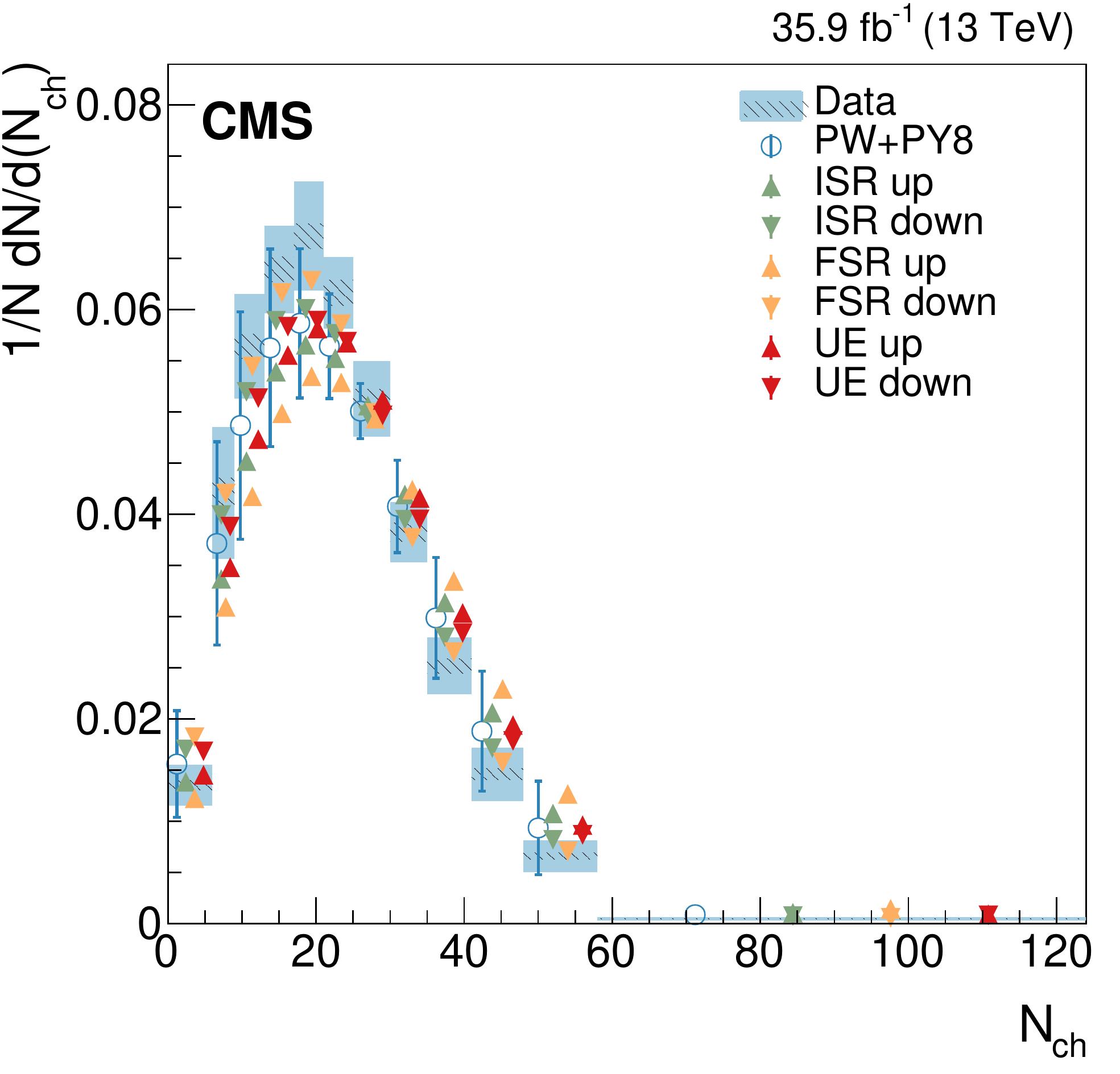}
\includegraphics[width=0.57\textwidth]{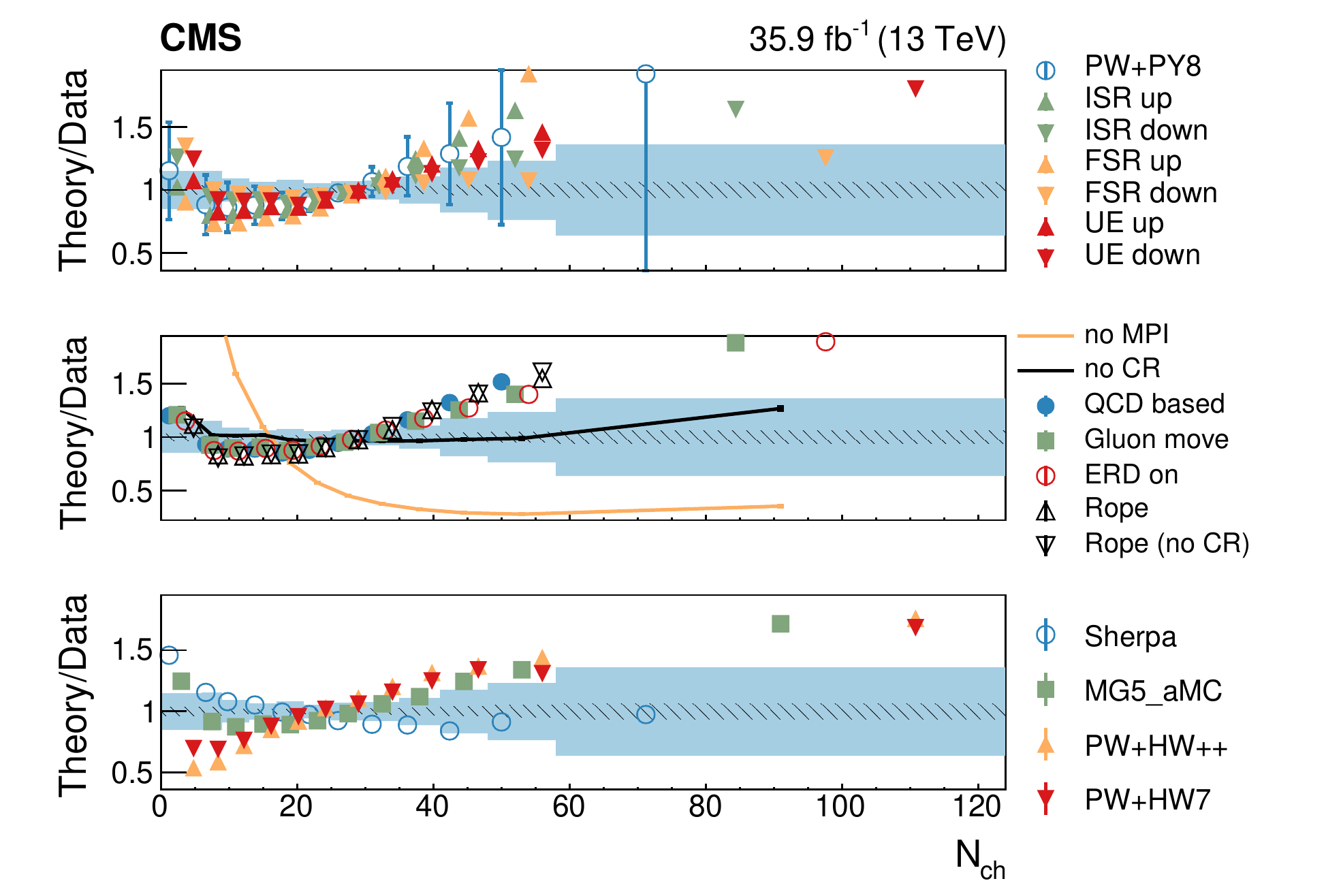}
\caption{The charged particle multiplicity of the UE in \ttbar events~\cite{TOP17015}.  The data are compared to several different MC generators.}
\label{fig:UE_Nch}
\end{figure}

Figure~\ref{fig:UE_Nch} shows the charged particle multiplicity of the UE, along with comparisons to different MC generators.  The predictions of the \POWHEGPYTHIA generator configuration are also shown after varying several parameters sensitive to the modelling of \ttbar production, such as the ISR and FSR scales, and the model used to simulate colour reconnection (CR).  The simulation of multiparton interactions (MPI) and CR are found to be important in achieving an accurate description of the measured distributions, demonstrated by the effect of turning off the simulation of these processes.  The data also favours a lower value of the FSR scale, \alpsfsr, used in \PYTHIA.

\begin{figure}[htb]
\centering
\includegraphics[width=0.47\textwidth]{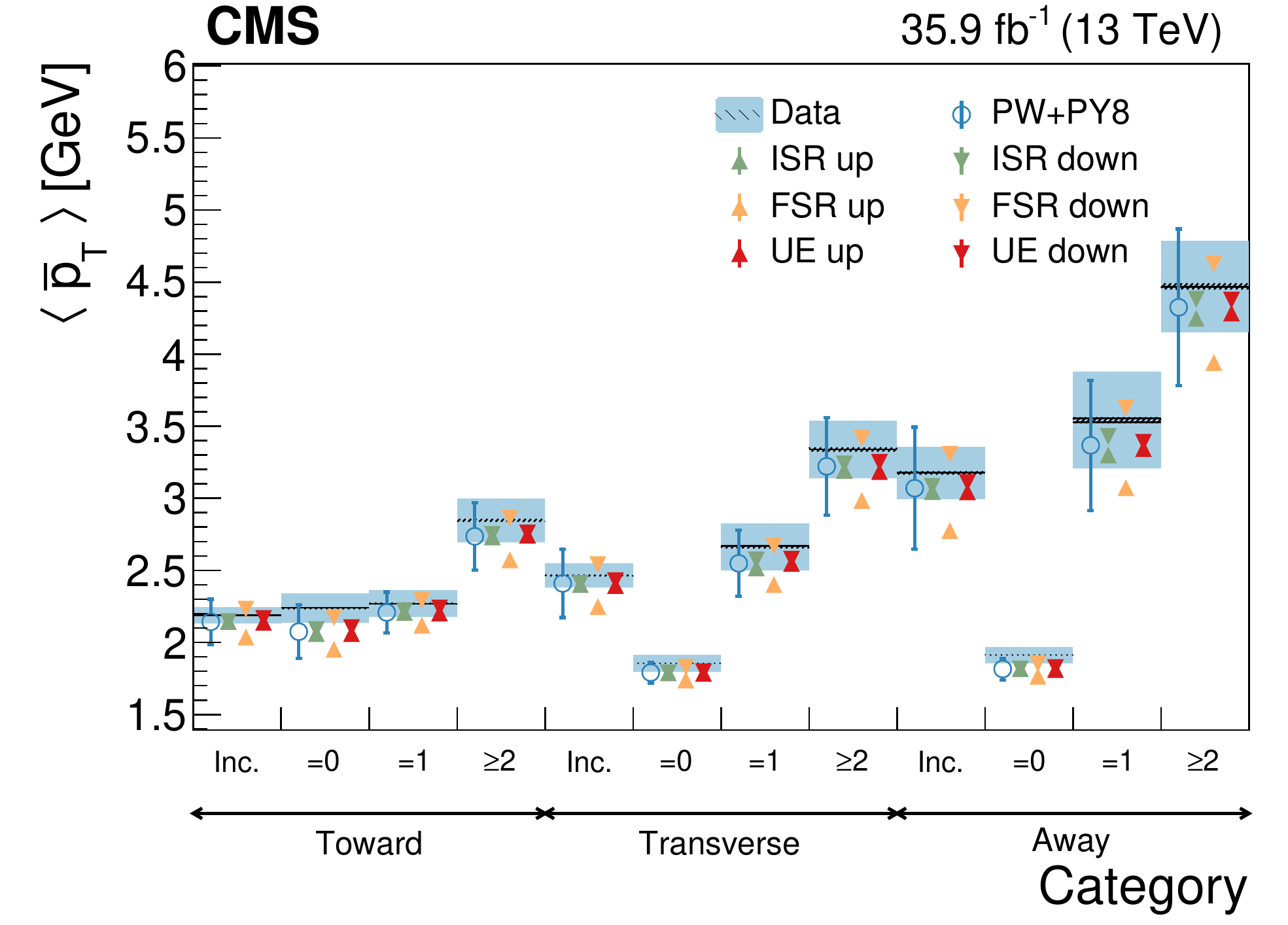}
\includegraphics[width=0.47\textwidth]{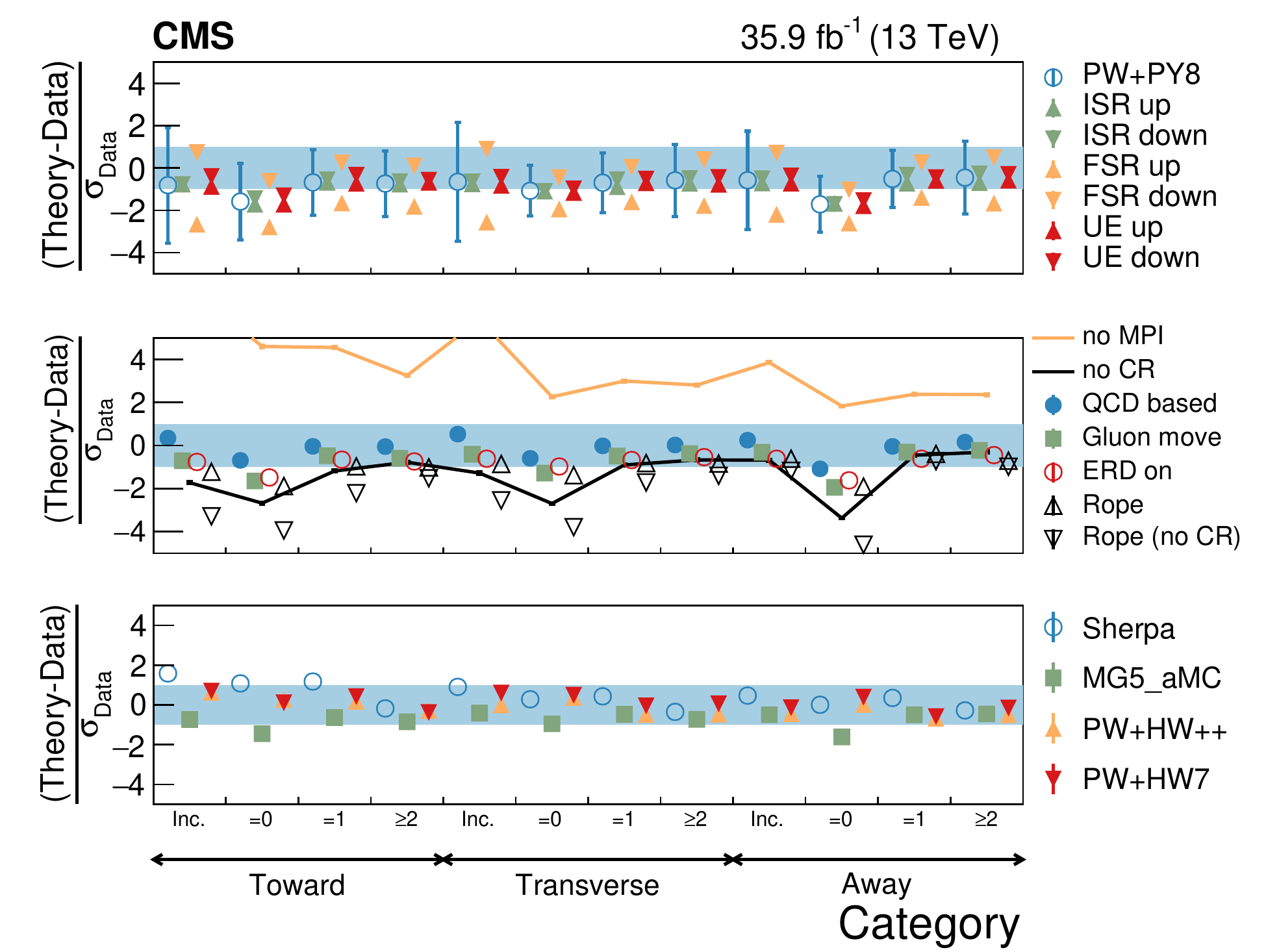}
\caption{The mean \ptbar of the UE in \ttbar events with different jet multiplicity, in the "towards", "transverse", and "away" regions~\cite{TOP17015}.  The data are compared to several different MC generators.}
\label{fig:UE_ptbar}
\end{figure}

In order to enhance the sensitivity of the measured distributions to different aspects of the simulation, the data are also categorised into regions of different jet multiplicity, dilepton transverse momentum (\pt), and dilepton invariant mass (\llmass). 
The difference of the azimuthal angle, \deltaPhi, between a charged particle in the UE and the dilepton momentum vector, \ptllvec, 
is also used to define regions in the transverse plane.  
These are the "towards" ($\deltaPhi<\ang{60}$), "transverse" ($\ang{120}<\deltaPhi<\ang{60}$), and "away" regions ($\deltaPhi>\ang{120}$).  
The mean \pt of charged particles, \ptbar, in the UE, as a function of the jet multiplicity in these different regions is shown in Figure~\ref{fig:UE_ptbar}.  In each bin of jet multiplicity, the mean of the \ptbar distribution in that bin is shown.

\section{Jet Substructure using \ttbar~events}
\label{sec:jet}
\begin{figure}[htb]
\centering
\includegraphics[width=0.45\textwidth]{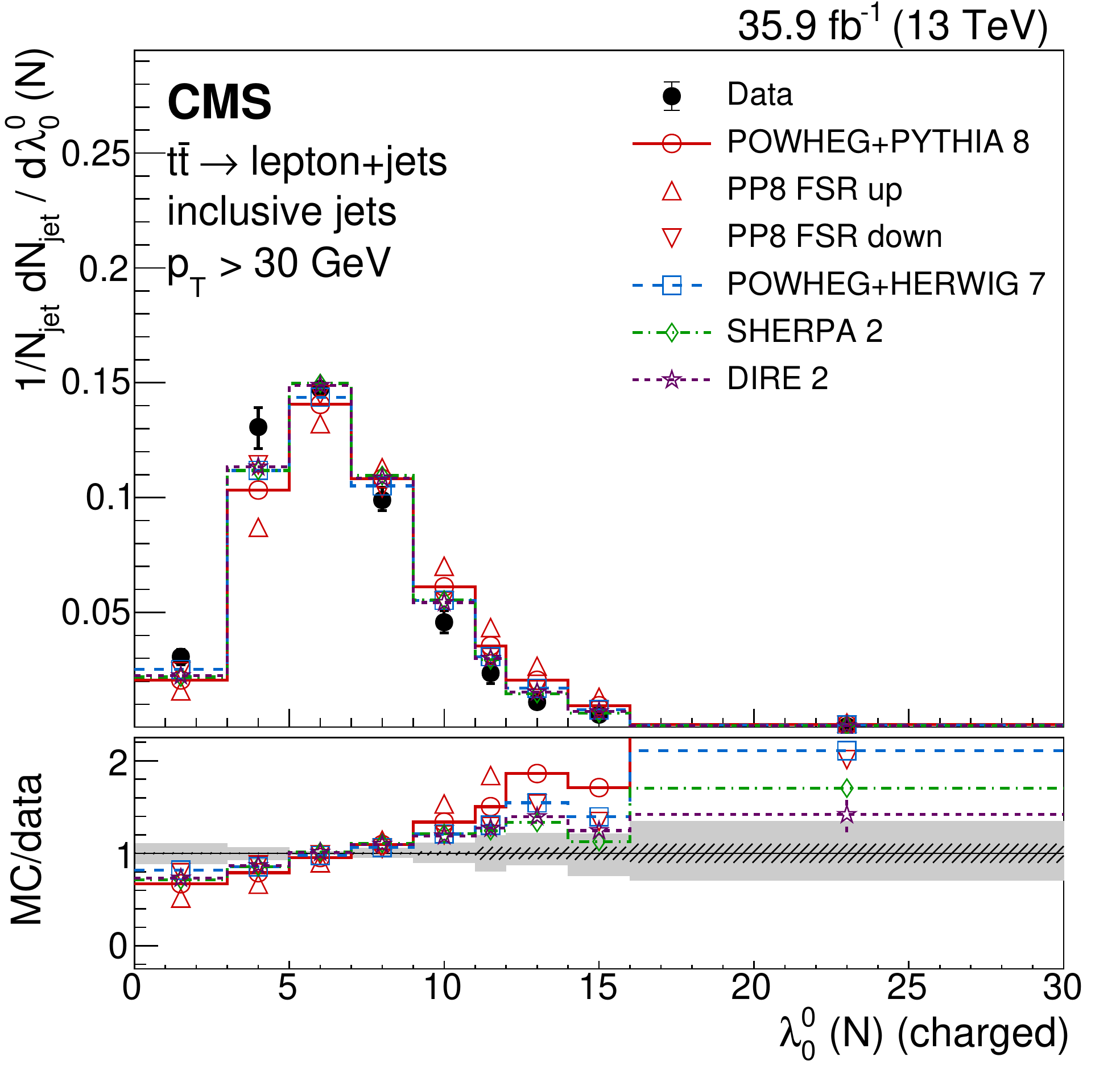}
\includegraphics[width=0.45\textwidth]{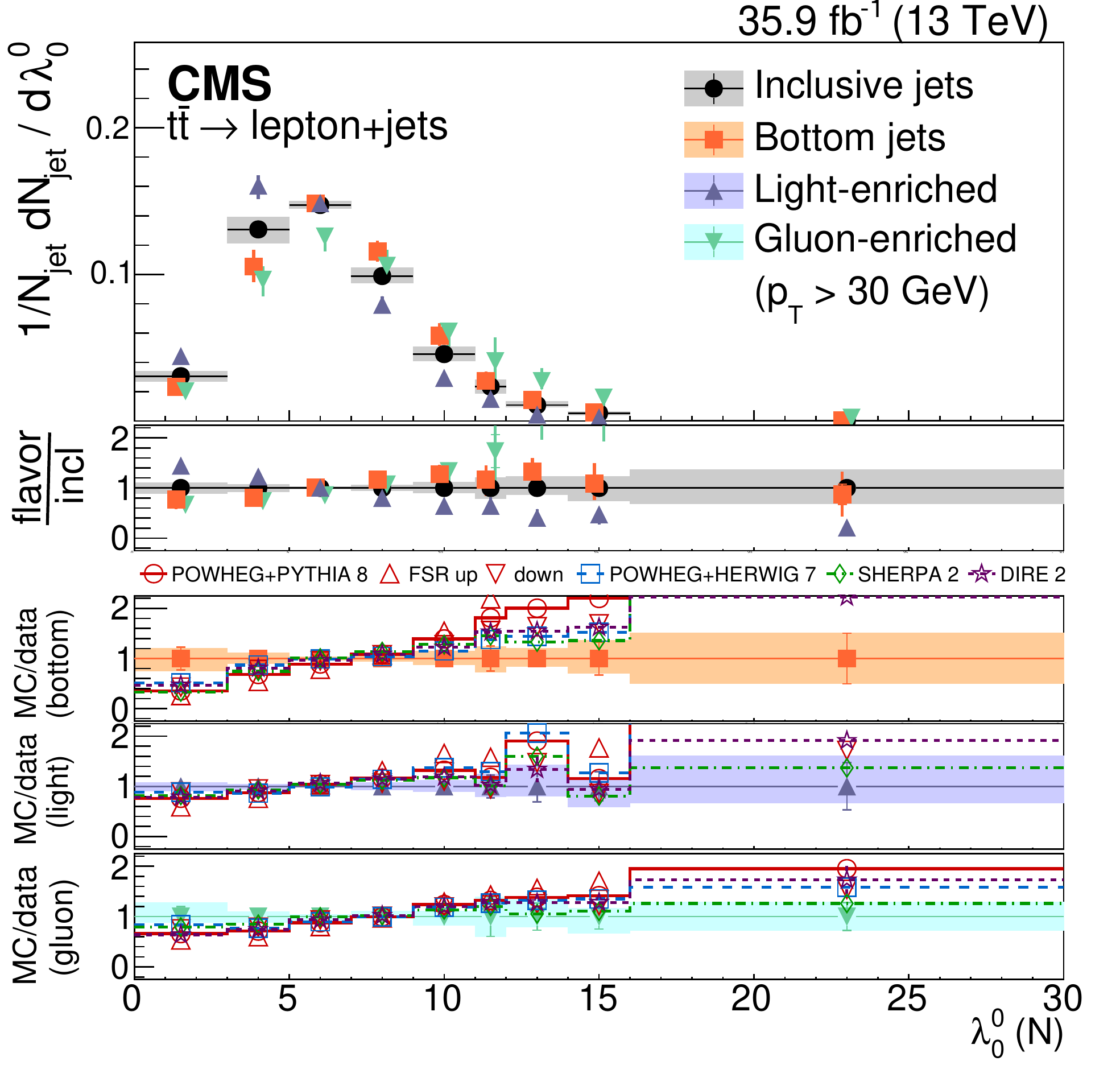}
\caption{The charged particle multiplicity in jets~\cite{TOP17013}.  On the left, the data are compared to different MC generators.  On the right, the distributions are shown for data samples that are enriched in {\bJet}s, light jets, and gluon jets.}
\label{fig:Jet_Nch}
\end{figure}

Top quark pair events with one lepton in the final state provide an abundance of jets.  The substructure of these jets has been studied by CMS using 35.9\fbinv of data at $\sqrts=13\TeV$~\cite{TOP17013}, testing the applicability of the PS models in the LHC environment.  Several families of observables are measured, such as generalised angularities, energy correlation functions, and observables related to the soft drop algorithm. The charged particle multiplicity inside the jets, unfolded to particle level in a phase space similar to that used at reconstruction level, is shown in Figure~\ref{fig:Jet_Nch}.  Samples of jets enriched in bottom quark, light quark, and gluon jets are also obtained, and the charged particle multiplicity for these different samples of jets is also shown in Figure~\ref{fig:Jet_Nch}.  The distributions measured in the analysis typically show that the data again favours a lower value of \alpsfsr in \PYTHIA.  In addition to this, the agreement of the different predictions with the data is better for the light quark and gluon jet enriched samples, and worse for the \bJet enriched sample.

\section{Extraction of \alpsfsr}

Both measurements presented in Sections~\ref{sec:UE} and~\ref{sec:jet} observe that a lower value of \alpsfsr is favoured by the data, compared to that used in \cuettune~\PYTHIA tune.  As a consequence of this, both measurements extract a value for \alpsfsr by performing a scan of the \chis between the data and predictions from \POWHEGPYTHIA as a function of \alpsfsr.  The most sensitive variable from the former analysis was found to be \ptbar, and the extracted value was $\alpsfsr=0.120^{+0.006}_{-0.006}$.  In the latter analysis, the angle between groomed subjets, \deltaRg, which is related to the jet width, was found to be the most sensitive variable, and the value extracted was $\alpsfsr=0.130^{+0.016}_{-0.020}$.
Both extracted values are somewhat lower than that in the $\cuettune$ tune ($\alpsfsr=0.1365$).  In addition to this, the uncertainties in the extracted value from the \ptbar distribution are approximately equivalent to a $\sqrt{2}$ variation in the renormalization scale.  This is smaller than the factor 2 variation typically considered by top quark measurements performed by CMS when estimating an uncertainty in this source.



\section{New \textsc{pythia 8} tunes}
A set of \PYTHIA tunes were derived by CMS where the same PDF set and \alpS value are used in all aspects of the simulation, such as the hard process, ISR, FSR, and the PS~\cite{CP5}.  The tunes were derived using UE and MB data, and the performance was tested with top data, along with UE, MB, and DY data.

\begin{figure}[htb]
\centering
\includegraphics[width=0.45\textwidth]{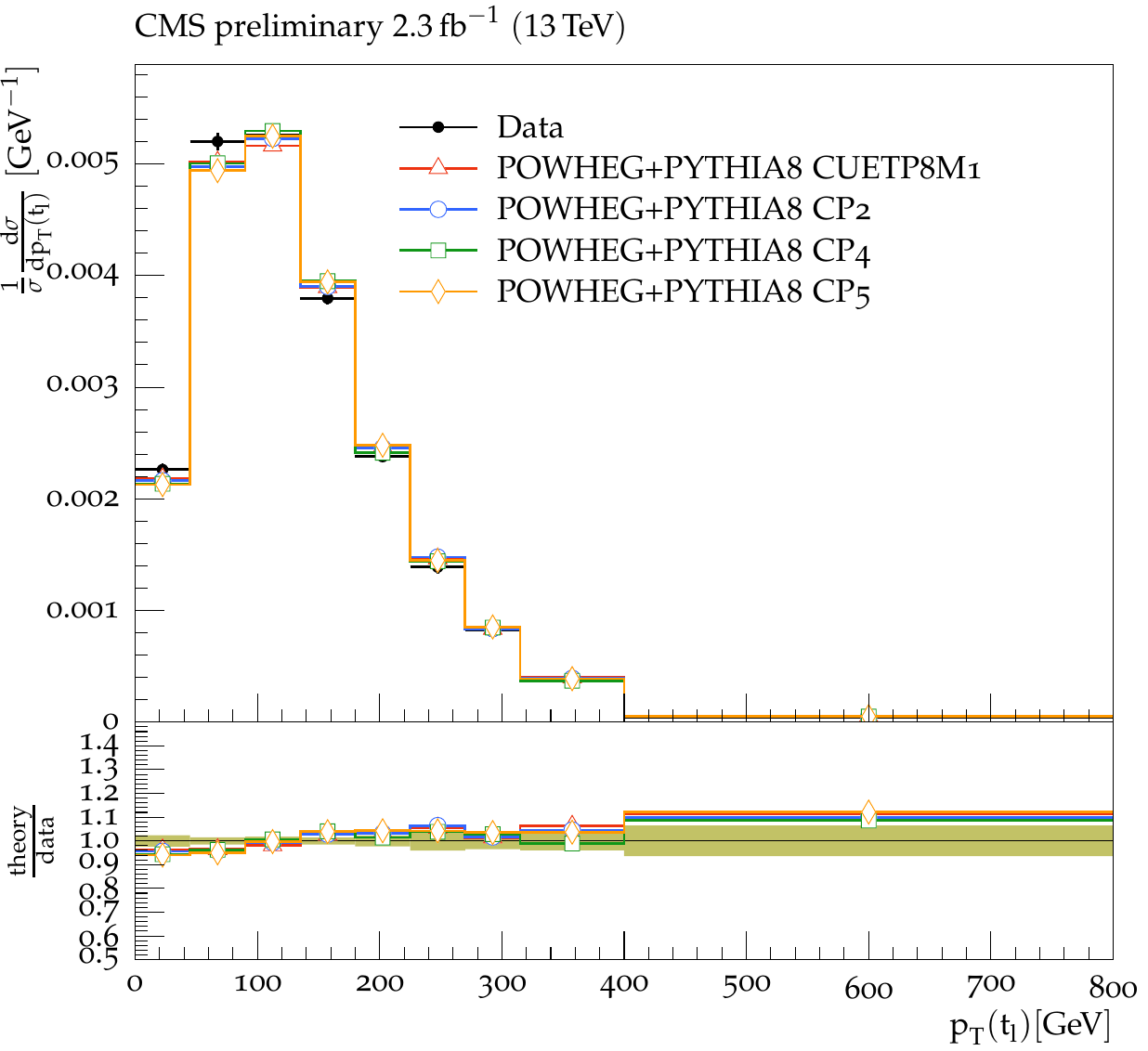}
\includegraphics[width=0.45\textwidth]{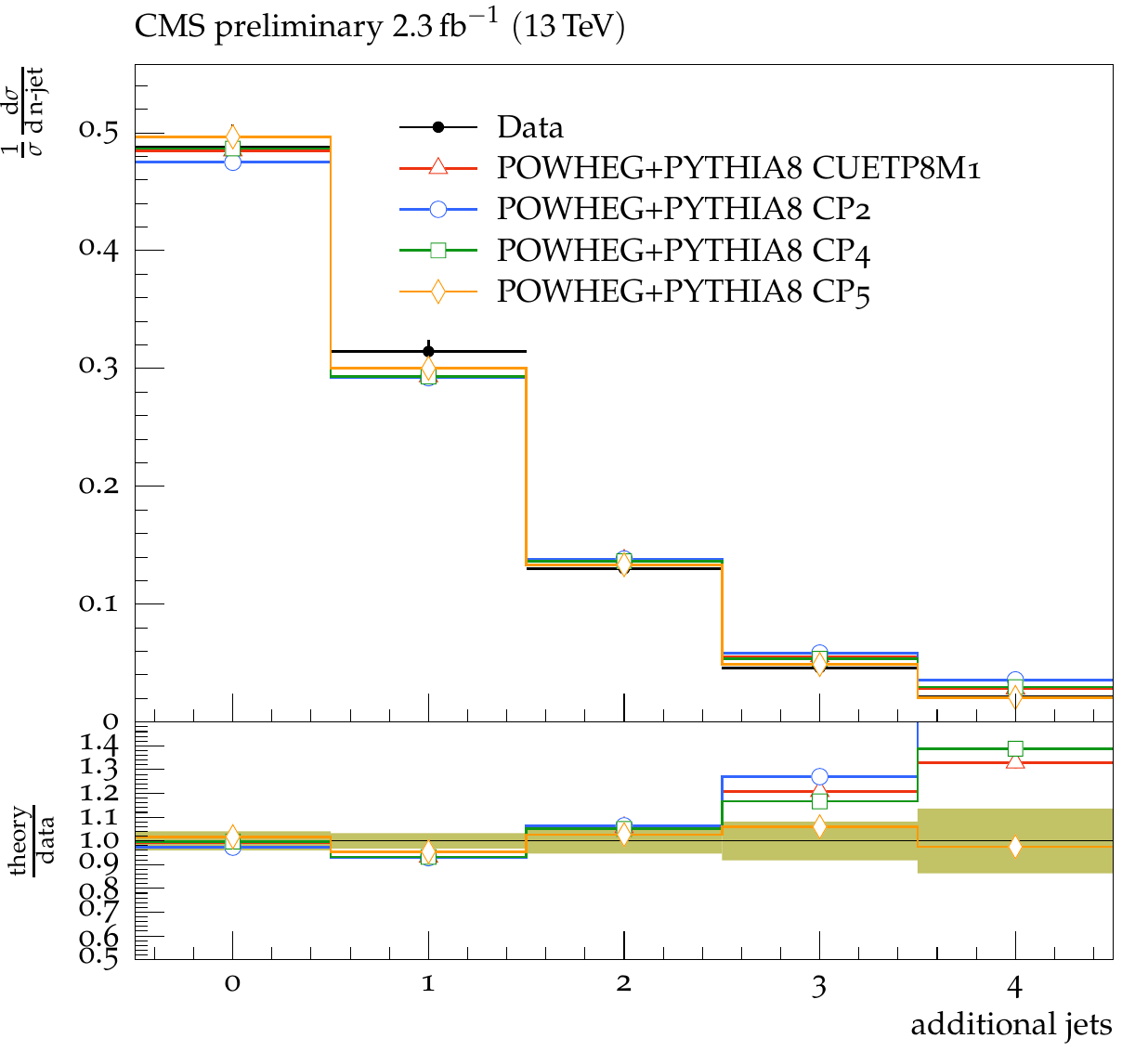}
\caption{The \pt of the leptonically decaying top quark (left) and jet multiplicity (right) in \ttbar events~\cite{CP5}.  The data are compared to predictions from \POWHEGPYTHIA, with various \PYTHIA tunes.}
\label{fig:CP}
\end{figure}

Figure~\ref{fig:CP} compares the predictions for the leptonically decaying top quark \pt and jet multiplicity from \POWHEGPYTHIA using the \oldcuettune, CP2, CP4, and CP5 tunes, with data.  CP2 uses the NNPDF3.1 LO PDF set~\cite{NNPDF}, and $\alpS=0.130$.  Both CP4 and CP5 uses NNPDF3.1 NNLO PDF set with $\alpS=0.118$, however CP5 also enables rapidity ordering for ISR.  The predictions from each setup are found to be able to describe the top quark kinematics, however only CP5 provides an accurate description for the whole range of additional jet multiplicity.

\begin{thebibliography}{99}

\bibitem{CMS} 
  CMS Collaboration,
  JINST {\bf 3}, S08004 (2008).


\bibitem{Pythia} 
  T.~Sjostrand, S.~Mrenna and P.~Z.~Skands,
  Comput.\ Phys.\ Commun.\  {\bf 178}, 852 (2008).
  
\bibitem{Powheg1} 
  P.~Nason,
  JHEP {\bf 0411}, 040 (2004).
  
\bibitem{Powheg2} 
  S.~Frixione, P.~Nason and C.~Oleari,
  JHEP {\bf 0711}, 070 (2007).

\bibitem{Powheg3} 
  S.~Alioli, P.~Nason, C.~Oleari and E.~Re,
  JHEP {\bf 1006}, 043 (2010).

\bibitem{PowhegHVQ} 
  S.~Frixione, M.~L.~Mangano, P.~Nason and G.~Ridolfi,
  Adv.\ Ser.\ Direct.\ High Energy Phys.\  {\bf 15}, 609 (1998).

\bibitem{CUETP8M2T4} 
  CMS Collaboration,
  CMS-PAS-TOP-16-021 (2016),\\
  http://cds.cern.ch/record/2235192.
 
\bibitem{CUETP8M1} 
  CMS Collaboration,
  Eur.\ Phys.\ J.\ C {\bf 76}, no. 3, 155 (2016).

\bibitem{MG} 
  J.~Alwall {et al.},
  JHEP {\bf 1407}, 079 (2014).

\bibitem{Sherpa} 
  T.~Gleisberg, S.~Hoeche, F.~Krauss, M.~Schonherr, S.~Schumann, F.~Siegert and J.~Winter,
  JHEP {\bf 0902}, 007 (2009).

\bibitem{Herwig7_1} 
  M.~Bahr {et al.},
  Eur.\ Phys.\ J.\ C {\bf 58}, 639 (2008).

\bibitem{Herwig7_2} 
  J.~Bellm {et al.},
  Eur.\ Phys.\ J.\ C {\bf 76}, no. 4, 196 (2016).
 
\bibitem{TOP17015} 
  CMS Collaboration,
  arXiv:1807.02810 (Submitted to EPJC).

\bibitem{TOP17013} 
  CMS Collaboration,
  Phys.\ Rev.\ D {\bf 98}, no. 9, 092014 (2018)

\bibitem{CP5} 
  CMS Collaboration,
  CMS-PAS-GEN-17-001 (2018),\\
  http://cds.cern.ch/record/2636284.

\bibitem{NNPDF} 
  R.~D.~Ball {et al.} (NNPDF Collaboration),
  Eur.\ Phys.\ J.\ C {\bf 77}, no. 10, 663 (2017).

\end{thebibliography}
\end{document}